\documentclass{article}

\usepackage{arxiv}

\usepackage{microtype}
\usepackage{graphicx}
\usepackage{subfigure}
\usepackage{booktabs}
\usepackage{comment}
\usepackage{wrapfig}
\usepackage{arydshln}
\usepackage{enumitem}
\usepackage{multirow}
\usepackage{url}
\usepackage{natbib}
\usepackage{authblk}

\usepackage{hyperref}

\usepackage{amsmath}
\usepackage{amssymb}
\usepackage{mathtools}
\usepackage{amsfonts}
\usepackage{bm}
\usepackage{bbm}

\usepackage{algorithm}
\usepackage{algorithmic}

\def\eqref#1{equation~\ref{#1}}

\def\1{\bm{1}}

\DeclareMathAlphabet{\mathsfit}{\encodingdefault}{\sfdefault}{m}{sl}
\SetMathAlphabet{\mathsfit}{bold}{\encodingdefault}{\sfdefault}{bx}{n}

\def\calW{{\mathcal{W}}}

\def\bbE{{\mathbb{E}}}

\usepackage{amsthm}

\theoremstyle{plain}

\newtheorem{theorem}{Theorem}[section]

\newtheorem{assumption}[theorem]{Assumption}

\usepackage[textsize=tiny]{todonotes}

\renewcommand{\eqref}[1]{(\ref{#1})}

\usepackage{xcolor}
\newcount\Comments  
\newcommand{\kibitz}[2]{\ifnum\Comments=1\textcolor{#1}{#2}\fi}

\newcommand*{\email}[1]{\texttt{#1}}

\usepackage[capitalize,noabbrev]{cleveref}

\allowdisplaybreaks

\title{A Note on Doubly Robust Estimator in Regression Discontinuity Designs}
\renewcommand{\shorttitle}{Doubly Robust Regression Discontinuity Designs}

\author{Masahiro Kato}
\affil{Department of Basic Science, The University of Tokyo\\
\email{mkato-csecon@g.ecc.u-tokyo.ac.jp}}

\begin{document}

\maketitle

\begin{abstract}
This note introduces a doubly robust (DR) estimator for regression discontinuity (RD) designs. RD designs provide a quasi-experimental framework for estimating treatment effects, where treatment assignment depends on whether a running variable surpasses a predefined cutoff. A common approach in RD estimation is the use of nonparametric regression methods, such as local linear regression. However, the validity of these methods still relies on the consistency of the nonparametric estimators. In this study, we propose the DR-RD estimator, which combines two distinct estimators for the conditional expected outcomes. The primary advantage of the DR-RD estimator lies in its ability to ensure the consistency of the treatment effect estimation as long as at least one of the two estimators is consistent. Consequently, our DR-RD estimator enhances robustness of treatment effect estimators in RD designs.
\end{abstract}

\section{Introduction}
Regression discontinuity (RD) designs provide a framework for natural experiments, enabling treatment effect estimation when treatment assignment is determined by whether a running variable crosses a fixed threshold, commonly referred to as a cutoff \cite{ImbensWooldridge2009}. One notable advantage of RD designs is that they do not require overlap in the support of the treatment and control groups for identification. This flexibility has led to the widespread application of RD designs across various empirical fields, including economics, political science, and epidemiology \citep{Klaauw2002,Black1999,AngristLevy1999,Lee2008randomized}.

\subsection{Problem Formulation}
This study focuses on sharp RD designs, where treatment assignment is deterministically determined by the value of the running variable.

We begin by outlining the problem setting within the Neyman-Rubin causal model \citep{Neyman1923,Rubin1974}. Let $d \in \{1, 0\}$ represent a binary treatment (or intervention). For each $d \in \{1, 0\}$, let $Y(d) \in \mathcal{Y} \subset \mathbb{R}$ denote a potential outcome, where the pair $(Y(1), Y(0))$ follows a distribution $P_0$, and $\mathcal{Y}$ represents the outcome space. For simplicity, we assume $Y(d)$ is a bounded random variable. Our primary interest is in estimating the treatment effect, defined as the difference $Y(1) - Y(0)$ \citep{ImbensRubin_2015}.

Next, we define the observations. We refer to individuals, firms, or other entities receiving the treatment as units. Suppose there are $n$ units, indexed as $1, 2, \dots, n$. For each unit $i \in \{1, 2, \dots, n\}$, let $D_i \in \{1, 0\}$ denote the assigned treatment, and let $Y_i \in \mathcal{Y}$ represent the observed outcome, defined by
\begin{align*}
    Y_i \coloneqq \mathbbm{1}[D_i = 1] Y_i(1) + \mathbbm{1}[D_i = 0] Y_i(0),
\end{align*}
where $(Y_i(1), Y_i(0))$ is an independent and identically distributed (i.i.d.) random variable drawn from the same distribution as $(Y(1), Y(0))$. 

Additionally, let 
\[
X_i = \begin{pmatrix} W_i \\ Z_i \end{pmatrix} \in \mathcal{X} \subset \mathbb{R}^d
\]
represent $d$-dimensional covariates (where $d \geq 1$), with $W_i \in \mathcal{W} \subset \mathbb{R}$ as the running variable and $Z_i \in \mathcal{Z} \subset \mathbb{R}^{d-1}$ as additional covariates. The spaces $\mathcal{X}$, $\mathcal{W}$, and $\mathcal{Z}$ correspond to the domains of $X$, $W$, and $Z$, respectively. In cases where $X_i = W_i$, we consider $d = 1$. Our observations thus consist of $\{(Y_i, D_i, X_i)\}_{i=1}^n$.

In a sharp RD design, treatment assignment $D_i$ is a deterministic function of the running variable $W_i$, defined by:
\begin{align*}
    D_i = \mathbbm{1}\big[ W_i \geq c \big],
\end{align*}
where $c \in \calW$ denotes the cutoff or threshold value.

Our objective is to estimate the average treatment effect (dTE) at the cutoff, defined as
\[
\tau_0 \coloneqq \mathbb{E}\left[Y(1) - Y(0) \mid W = c\right],
\]
commonly referred to as the conditional average treatment effect (CATE) in the causal inference literature.

In RD designs, this parameter is identifiable under the following assumption\footnote{A more general assumption is that for all $y \in \mathcal{Y}$ and each $d \in \{1, 0\}$, the conditional distribution function $F_d(y, W)$ is continuous in $w$, where $F_d(y, W)$ is the conditional cumulative distribution function of $Y(d)$.}.

\begin{assumption}[Continuity of Conditional Regression Functions]
\label{asm:continuity}
    For all $y \in \mathcal{Y}$ and each $d \in \{1, 0\}$, the conditional outcome function $\mu(d, W)$ is continuous in $w$.
\end{assumption}

With this assumption, we have 
\[
\mathbb{E}\left[Y(1) \mid W = c\right] = \lim_{w \uparrow c} \mathbb{E}\left[Y(1) \mid W = w\right],\quad 
\mathbb{E}\left[Y(0) \mid W = c\right] = \lim_{w \downarrow c} \mathbb{E}\left[Y(0) \mid W = w\right],
\]
leading to
\[
\tau_0 = \lim_{w \uparrow c} \mathbb{E}\left[Y(1) \mid W = w\right] - \lim_{w \downarrow c} \mathbb{E}\left[Y(0) \mid W = w\right].
\]

The RD design has several advantages. First, it offers interpretability. Second, it does not require the common support assumption, unlike conventional treatment effect estimation, which typically assumes that for all $x$, the treatment assignment probability $p(d \mid x)$ satisfies $p(d \mid x) \in (0, 1)$. In sharp RD designs, $p(d \mid x) = 0$ holds for $x = (w, c)$ when $w \leq c$, and $p(d \mid x) = 1$ when $w > c$.

Due to the nature of the RD setting, only the conditional average treatment effect (CATE) $\tau(c)$ around the cutoff $c$ is identifiable. Without additional assumptions, the average treatment effect $\mathbb{E}\left[\tau(X)\right]$ and the CATE $\tau(x)$ for all $x \in \mathcal{X}$ are not identifiable.

\subsection{Existing RD Methods}
Existing studies estimate treatment effects using parametric or nonparametric regression methods, including kernel (Nadaraya-Watson) regression, local linear regression, and sieve regression. Among these, local linear regression is widely recommended \citep{ImbensLemieux2008} because it reduces bias compared to kernel regression—a key advantage in RD settings where the focus is on the conditional expected outcome at the boundary (i.e., the cutoff point). Since RD designs require estimation only at this boundary, there is no need to estimate the functional form across the entire range of $x$, as would be necessary in sieve regression. Consequently, local linear regression is frequently preferred in RD applications.

\subsection{Our Contributions}
This study develops a doubly robust (DR) estimator for the treatment effect in sharp RD designs, incorporating estimators of both the localization function and conditional outcomes. The estimator possesses the double-robustness property, meaning that if either the localization function estimator or the conditional outcome estimator is consistent, the treatment effect estimator remains consistent. 

The doubly robust structure of our estimator enables flexible modeling of the treatment effect. In RD designs, estimation approaches are typically divided into global and local methods, with the former represented by sieve regression and the latter by local linear regression. In traditional approaches, one must choose between these methods. However, our estimator integrates both approaches, allowing the localization function estimator to align with the local approach while the conditional outcomes are estimated using the global approach.

Note that DR estimators in causal inference are typically proposed in settings where overlapping support exists \citep{BangRobins2005}, which differs from the setting in RD designs. Our method enables the construction of a DR estimator without assuming such a common support.

\subsection{Organization}
In Section~\ref{sec:dr-rd}, we define our DR-RD estimator and present the main theorem regarding its asymptotic normality. Section~\ref{sec:consistency} explains the doubly robust property of the DR-RD estimator. In Section~\ref{sec:est_nuisance}, we show examples of implementations of the DR-RD estimator.

\paragraph{Notation:} Our notation primarily follows \citet{ImbensLemieux2008}, with certain modifications for clarity.

\section{The DR-RD Estimator}
\label{sec:dr-rd}
This section defines our proposed DR-RD estimator. Let us denote the conditional expected outcome by $\mu_0(d\mid W, Z) = \bbE\left[Y(d) \mid W, Z\right]$. The DR-RD estimator consists of the following two-stage procedure. In the first stage, we estimate $\mu_0$. In the second stage, we estimate the conditional expected residual of the first-stage estimator.

\paragraph{First stage.}
Let $\widehat{\mu}_n$ be some estimator of $\mu_0$. We can use various methods for this estimation, such as Nadaraya-Watson regression, local linear regression, series regression, and regression using neural networks. Examples are provided in Section~\ref{sec:est_nuisance}.

\paragraph{Second stage.}
In the second stage, we estimate the conditional expected residual of the first-stage estimator using Nadaraya-Watson regression and obtain the treatment effect estimator. 

We define
\begin{align}
\label{eq:kernel}
    \widehat{r}_{n, c}(1, W) &= \frac{1}{\frac{1}{\sum_{i=1}^n \mathbbm{1}[W_i \geq c]}\sum_{i: W_i \geq c} K\left(\frac{W_i - c}{h}\right)} K\left(\frac{W - c}{h}\right),\\
    \widehat{r}_{n, c}(0, W) &= \frac{1}{\frac{1}{\sum_{i=1}^n \mathbbm{1}[W_i < c]}\sum_{i: W_i < c} K\left(\frac{W_i - c}{h}\right)} K\left(\frac{W - c}{h}\right).\nonumber
\end{align}

Then, we define the DR-RD estimator of $\tau(c)$ as
\begin{align*}
    \widehat{\tau}^{\mathrm{DR}\mathchar`-\mathrm{RD}} \coloneqq \frac{1}{n} \sum_{i=1}^n \Big(\psi(1, Y_i, W_i, Z_i; \widehat{\mu}_n, \widehat{r}_{n, c}) - \psi(0, Y_i, W_i, Z_i; \widehat{\mu}_n, \widehat{r}_{n, c})\Big),
\end{align*}
where $\psi$ is defined as
\begin{align*}
    \psi(d, Y_i, W_i, Z_i; \widehat{\mu}_n, \widehat{r}_{n, c}) \coloneqq \mathbbm{1}\left[D_i = d\right] \Big(Y_i - \widehat{\mu}_n(d \mid W_i, Z_i)\Big) \widehat{r}_{n, c}(d, W_i) + \widehat{\mu}_n(d \mid c, Z_i).
\end{align*}

Here,
$
\frac{1}{n} \sum_{i=1}^n \mathbbm{1}\left[D_i = d\right] \Big(Y_i - \widehat{\mu}_n(d \mid W_i, Z_i)\Big) \widehat{r}_{n, c}(d, W_i)
$
corresponds to an estimator of the expected residual $Y_i(d) - \widehat{\mu}_n(d, W_i, Z_i)$ conditioned on $W_i$. This second-stage estimator corrects the bias in the firs-stage estimator. Note that the second-stage estimator is also interpreted as an estimator of the estimation error $\mu_0(d\mid W_i, Z_i) - \widehat{\mu}_n(d\mid W_i, Z_i)$. 

\section{Consistency and Double Robustness}
\label{sec:consistency}
Our proposed DR-RD estimator has the property of double robustness; that is, if either $\mu_0$ or the conditional expected residual is consistently estimated, we can consistently estimate $\tau_0$. This double robustness is a key motivation for our proposal of the DR-RD estimator.

Suppose that as $n \to \infty$, $\widehat{\mu}_n \xrightarrow{\mathrm{p}} \mu^\dagger$, 
\[
\frac{1}{\sum_{i=1}^n \mathbbm{1}[D_i = d]}\sum_{i: D_i = d} K\left(\frac{W_i - c}{h}\right) \xrightarrow{\mathrm{p}} \zeta^\dagger(d, c),
\]
and
\[
\frac{1}{n} \sum_{i=1}^n \mathbbm{1}\left[D_i = d\right] \left(Y_i - \widehat{\mu}_n(d \mid W_i, Z_i)\right) \widehat{r}_{n, c}(d, W_i) \xrightarrow{\mathrm{p}} \eta^\dagger(d, c).
\]
Additionally, suppose that if either $\mu^\dagger = \mu_0$ or (given $\mathrm{plim}_{n\to\infty}\widehat{\mu}_n = \mu^\dagger$)
\begin{align}
\label{eq:cond_conv}
    \eta^\dagger(d, c) = \mathbb{E}\left[Y(d) - \mu^\dagger(d\mid W, Z)\mid W = c\right] = \mathbb{E}\left[\mu_0(d\mid W, Z) - \mu^\dagger(d\mid W, Z)\mid W = c\right]
\end{align}
holds, then $\widehat{\tau}^{\mathrm{DR}\mathchar`-\mathrm{RD}} \xrightarrow{\mathrm{p}} \tau_0$ holds as $n \to \infty$.

First, we show that if $\mu^\dagger = \mu_0$, then 
$\widehat{\tau}^{\mathrm{DR}\mathchar`-\mathrm{RD}} \xrightarrow{\mathrm{p}} \tau_0$ holds as $n \to \infty$. This is because for each $d \in \{1, 0\}$, it holds that
\begin{align*}
    &\mathbb{E}\left[\mathbbm{1}\left[D = d\right] \Big(Y - \mu_0(d \mid W, Z)\Big) \frac{K\left(\frac{W - c}{h}\right)}{\zeta^\dagger(d, c)} + \mu_0(d \mid c, Z)\right] = \mathbb{E}\left[\mu_0(d \mid c, Z)\right].
\end{align*}
Then, from the law of large numbers, $\widehat{\tau}^{\mathrm{DR}\mathchar`-\mathrm{RD}} \xrightarrow{\mathrm{p}} \tau_0$ holds. 

Similarly, we can show that if \eqref{eq:cond_conv} holds, then $\widehat{\tau}^{\mathrm{DR}\mathchar`-\mathrm{RD}} \xrightarrow{\mathrm{p}} \tau_0$ holds as $n \to \infty$. This is because for each $d \in \{1, 0\}$, it holds that
\begin{align*}
    &\mathbb{E}\left[\mu_0(d\mid W, Z) - \mu^\dagger(d\mid W, Z)\mid W = c\right] + \mathbb{E}\left[\mu^\dagger(d \mid c, Z)\right] = \mathbb{E}\left[\mu_0(d \mid c, Z)\right].
\end{align*}
Then, from the law of large numbers, $\widehat{\tau}^{\mathrm{DR}\mathchar`-\mathrm{RD}} \xrightarrow{\mathrm{p}} \tau_0$ holds. 

Based on these results, we have the following theorem on consistency.

\begin{theorem}[Consistency and Double Robustness] 
Suppose that as $n\to\infty$, $\widehat{\mu}_n \xrightarrow{\mathrm{p}} \mu^\dagger$, 
\[
\frac{1}{\sum_{i=1}^n \mathbbm{1}[D_i = d]}\sum_{i: D_i = d} K\left(\frac{W_i - c}{h}\right) \xrightarrow{\mathrm{p}} \zeta^\dagger(d, c),
\]
and
\[
\frac{1}{n} \sum_{i=1}^n \mathbbm{1}\left[D_i = d\right] \Big(Y_i - \widehat{\mu}_n(d \mid W_i, Z_i)\Big) \widehat{r}_{n, c}(d, W_i) \xrightarrow{\mathrm{p}} \eta^\dagger(d, c)
\]
hold, where $\mu^\dagger$, $\zeta^\dagger$, and $\eta^\dagger$ are constant functions or parameters independent of $n$. Additionally, suppose that either $\mu^\dagger = \mu_0$ or 
\[
\eta^\dagger(d, c) = \mathbb{E}\left[\mu_0(d\mid W, Z) - \mu^\dagger(d\mid W, Z)\mid W = c\right]
\]
holds for each $d \in \{1, 0\}$. If Assumption~\ref{asm:continuity} holds, then we have
\[
\widehat{\tau}^{\mathrm{DR}\mathchar`-\mathrm{RD}} \xrightarrow{\mathrm{p}} \tau_0 \quad (n\to \infty).
\]
\end{theorem}

\section{Examples of the DR-RD Estimators}
\label{sec:est_nuisance}
This section introduces examples of the DR-RD estimators. 

\subsection{Integration of Local and Global Approaches}
Under our estimator, we can integrate both local and global approaches in nonparametric regression. For example, as shown below, we can use a global approach, such as sieve regression, to estimate $\mu_0$, while we use local regression (Nadaraya-Watson regression) to estimate the conditional expected residual:
\begin{align*}
    \psi\left(d, Y_i, W_i, Z_i; \widehat{\mu}_n, \widehat{r}_{n, c}\right) \coloneqq 
    \underbrace{\mathbbm{1}\left[D_i = d\right] \Big(Y_i - \widehat{\mu}_n(d \mid W_i, Z_i)\Big) \widehat{r}_{n, c}(d, W_i)}_{\text{Local approach}} 
    + \underbrace{\widehat{\mu}_n(d \mid c, Z_i)}_{\text{Global approach}}.
\end{align*}

The local approach focuses on estimating $\mu_0(d\mid w, z)$ at the cutoff value $w = c$, while the global approach accounts for the overall shape of $\mu_0(d\mid w, z)$. If $\mu_0(d\mid w, z)$ exhibits a global pattern across all $(w, z)$, using a global approach may improve estimation accuracy.

\subsection{Integration of Linear and Non-Linear Estimators}
Another example is to use non-linear estimators for estimating $\mu_0$. Many traditional nonparametric estimators, such as Nadaraya-Watson and sieve estimators, fall into the category of linear estimators, which take the form $\sum_{i=1}^n g(x; X_1, \dots, X_n) Y_i$, where $g$ is a weight function that depends only on the observed covariates $X_1, \dots, X_n$ and the target value $x$. In contrast, some machine learning models, such as neural networks, are non-linear estimators.

It has been shown that neural networks can achieve better nonparametric estimation errors compared to linear estimators. For example, \citet{ImaizumiFukumizu2019} demonstrate that when the true regression function belongs to the piecewise Hölder class, nonparametric regression with neural networks achieves smaller estimation errors than the minimax rate for linear estimators. Similarly, \citet{Suzuki2018Adaptivity} show that when the true regression function lies in the Besov space, neural networks can mitigate the curse of dimensionality.

Based on these findings, there are advantages to using non-linear estimators. Here, we construct the DR-RD estimator using non-linear estimators, such as neural networks, to estimate $\mu_0$, as follows:
\begin{align*}
    \psi\left(d, Y_i, W_i, Z_i; \widehat{\mu}_n, \widehat{r}_{n, c}\right) \coloneqq 
    \underbrace{\mathbbm{1}\left[D_i = d\right] \Big(Y_i - \widehat{\mu}_n(d \mid Z_i, W_i)\Big) \widehat{r}_{n, c}(d, W_i)}_{\text{Linear estimator}} 
    + \underbrace{\widehat{\mu}_n(d \mid c, Z_i)}_{\text{Non-linear estimator}}.
\end{align*}

\section{Conclusion}
In this study, we introduced a DR estimator for RD designs. Our DR-RD estimator is consistent if either the first-stage conditional expected outcome estimator or the second-stage conditional expected residual estimator is consistent. This double robustness property enhances the robustness of RD designs in empirical research. We also provided examples demonstrating the potential to reconcile local and global approaches in RD estimation, thereby broadening the methodological toolkit available to researchers working with complex data structures. Overall, our work contributes to the methodological advancement of RD designs, offering a flexible and robust framework.

\bibliography{arXiv.bbl}

\bibliographystyle{tmlr}

\end{document}